\documentclass[12pt,oneside]{article}
\usepackage{epsfig, cite}
\usepackage{color}
\usepackage{ifthen}
\usepackage{graphicx}

\def\p{\partial}

\def\half{{1\over 2}}
\def\({\left(}
\def\){\right)}
\def\[{\left[}
\def\]{\right]}

\def\e{\begin{equation}}
\def\q{\end{equation}}
\def\m{\begin{eqnarray}}
\def\n{\end{eqnarray}}

\def\cn{{\cal N}}

\begin{document}
\thispagestyle{empty} \setcounter{page}{0}
\renewcommand{\theequation}{\thesection.\arabic{equation}}

\vspace{2cm}

\begin{center}
{\huge N-vaton}

\vspace{1.4cm}

Qing-Guo Huang

\vspace{.2cm}

{\em School of physics, Korea Institute for
Advanced Study,} \\
{\em 207-43, Cheongryangri-Dong,
Dongdaemun-Gu, } \\
{\em Seoul 130-722, Korea}\\
\end{center}

\vspace{-.1cm}

\centerline{{\tt huangqg@kias.re.kr}} \vspace{1cm}
\centerline{ABSTRACT}
\begin{quote}
\vspace{.5cm}

In general there are a large number of light scalar fields in the
theories going beyond standard model, such as string theory, and
some of them can be taken as the candidates of curvatons. For
simplicity, we assume all of curvatons have the same decay rate and
suddenly decay into radiation at the same time. In order to
distinguish this scenario from the more general case, we call it
``N-vaton". We use $\delta {\cal N}$ formalism to calculate the
primordial power spectrum and bispectrum in N-vaton model and
investigate various bounds on the non-Gaussianity parameter
$f_{NL}$. A red tilted primordial power spectrum and a large value
of $f_{NL}$ can be naturally obtained if the curvature perturbation
generated by inflaton also makes a significant contribution to the
primordial power spectrum. As a realistic N-vaton model, we suppose
that the axions in the KKLT compactifications of Type IIB string
theory are taken as curvatons and a rich phenomenology is obtained.

\end{quote}
\baselineskip18pt

\noindent

\vspace{5mm}

\newpage

\setcounter{equation}{0}
\section{Introduction}

Most inflation models predict a nearly Gaussian distribution of the
primordial curvature perturbation. Deviations from an exactly
Gaussian distribution are characterized by a dimensionless parameter
$f_{NL}$ \cite{Bartolo:2004if}. In the case of single field
inflation model $f_{NL}\sim {\cal O}(n_s-1)$
\cite{Maldacena:2002vr}, which is constrained by WMAP
($n_s=0.960_{-0.013}^{+0.014}$) \cite{Komatsu:2008hk} to be much
less than unity. A Gaussian distribution of the primordial curvature
perturbation is still consistent with WMAP five-year data
\cite{Komatsu:2008hk}: \e -9<f_{NL}^{local}<111 \quad \hbox{and}
\quad -151<f_{NL}^{equil}<253 \quad (95\% \hbox{CL}), \q where
``local" and ``equil" denote the shapes of the non-Gaussianity. In
\cite{Yadav:2007yy} the authors reported that a positive large
non-Gaussianity \e 27<f_{NL}^{local}<147 \q is detected at $95\%$
C.L.. Planck is expected to bring the uncertainty of
$f_{NL}^{local}$ to be less than 5 \cite{Komatsu:2001rj}. If a large
value of $f_{NL}$ is confirmed by the forthcoming cosmological
observations, the simplest model of inflation is ruled out and some
very important new physics of the early Universe will be showed up.

In general a large number of light scalar fields are expected in the
theories beyond the standard model, such as string theory. The
consistent perturbative superstring theory can only live in
ten-dimensional spacetime. To connect string theory with
experiments, string theory must be compactified on some
six-dimensional manifold and many dynamical moduli fields emerges in
four dimensions. The typical number of moduli fields is $N\sim {\cal
O}(10^2-10^3)$. One can expect that the expectation values of some
of these scalar fields are displaced from the minimum of their
potential due to the quantum fluctuations during inflation. Usually
these scalar fields are subdominant during inflation and their
fluctuations are initially of isocurvature type. After the end of
inflation they are supposed to completely decay into thermalized
radiation before primordial nucleosynthesis and thus the
isocurvature perturbations generated by them are converted to be a
final adiabatic perturbations. These scalar fields are called
curvatons.

The curvaton mechanism to generate an initially adiabatic
perturbation deep in the radiation era is proposed in
\cite{Linde:1996gt,Enqvist:2001zp,Lyth:2001nq,Moroi:2001ct}. The
primordial curvature perturbation in curvaton model with single
curvaton has been discussed in
\cite{Linde:1996gt,Enqvist:2001zp,Lyth:2001nq,Moroi:2001ct,Lyth:2002my,Bartolo:2003jx,Malik:2006pm,Sasaki:2006kq}.
If many curvatons make contributions to the primordial density
perturbation, the calculation becomes much more complicated
\cite{Choi:2007fya,Assadullahi:2007uw,Valiviita:2008zb}. Since
curvaton model can give a large positive local-type non-Gaussianity,
recently some topics related to curvaton model are discussed in
\cite{Huang:2008ze,Ichikawa:2008iq,Multamaki:2008yv,Suyama:2008nt,Beltran:2008ei,Li:2008jn,Li:2008fm,Huang:2008qf}.

In single-curvaton model $f_{NL}$ is inverse proportional to the
fraction of curvaton energy density in the energy budget at the
epoch of curvaton decay. Smaller the energy density of curvaton,
larger the non-Gaussianity. Since the curvaton mass is smaller than
the Hubble parameter $H_*$ during inflation, or equivalently its
Compton wavelength is large compared to the curvature radius of the
de Sitter space $H_*^{-1}$, the gravitational effects play a crucial
role on the behavior of curvaton field in such a scenario. The
typical energy density of curvaton field in such a background is
roughly $H_*^4$, which leads to an upper bound on $f_{NL}$
\cite{Huang:2008ze}: $f_{NL}<522\cdot r^{1\over 4}$ (up to an order
one coefficient), where $r$ is the tensor-scalar ratio.

Since multiplicity of scalar fields is generally expected, we focus
on the multi-curvaton scenario in this paper. Here we consider a
special case in which all of curvatons have the same decay rate and
their masses are larger than the decay rate. In order to simplify
the calculation of the primordial curvature perturbation, we assume
all of curvatons suddenly decay into radiation at the same time. We
give a name, ``N-vaton", to this scenario. As a realistic N-vaton
model, the axions in the KKLT compactification of Type IIB string
theory are suggested to be curvatons. Based on the random matrix
theory, the mass spectrum of axion obeys the Marcenko-Pastur law and
a rich phenomenology of this model is shown up.

Our paper is organized as follows. In Sec. 2, we use the $\delta
\cn$ formalism \cite{Starobinsky:1986fxa,Sasaki:1995aw,Lyth:2005fi}
to calculate the primordial curvature perturbation on large scales
in N-vaton model. The various bounds on $f_{NL}$ are discussed in
Sec. 3. In Sec. 4 we consider a more general case where the
curvature perturbation generated by inflaton cannot be ignored and
we find that the spectral index of primordial power spectrum can be
red-tilted naturally. In Sec. 5, we propose a realistic N-vaton
model in which the axions in the KKLT compactification of Type IIB
string theory are taken as curvatons. At the end we give some
discussions on N-vaton model in Sec. 6.

\setcounter{equation}{0}
\section{Primordial curvature perturbation}

In this paper we consider that inflaton $\phi$ and curvatons
$\sigma_i$ are decoupled to each other. The action takes the form \e
S={M_p^2\over 2}\int d^4x\sqrt{-g} R + \int d^4x\sqrt{-g}
\[\half {\dot \phi}^2+\sum_{i=1}^N\half {\dot \sigma_i}^2-V(\phi,\sigma_i)\], \q
where $M_p=2.438\times 10^{18}$ GeV is the reduced Planck scale and
the potential $V(\phi,\sigma_i)$ is given by \e
V(\phi,\sigma_i)=V(\phi)+\half \sum_{i=1}^{N} m_i^2\sigma_i^2. \q
During inflation the total energy density is dominated by inflaton
potential $V(\phi)$ and the dynamics of the system is described by
the equation of motion of inflaton and the Friedmann equation: \m
\ddot \phi&+&3H\dot \phi+{dV(\phi)\over d\phi}=0, \\
H^2&\equiv&\({\dot a \over a}\)^2={1\over 3M_p^2}\(\half{\dot
\phi}^2+V(\phi)\). \n We also define some slow-roll parameters, such
as \e \epsilon={M_p^2\over 2}\({V'(\phi)\over V(\phi)}\)^2, \quad
\eta=M_p^2{V''(\phi)\over V(\phi)}. \q If $\epsilon\ll 1$ and
$|\eta|\ll 3$, inflaton slowly rolls down its potential.

In this paper we expand any field or perturbation at each order
$(n)$ as follows \e \zeta(t,{\bf x})=\zeta^{(1)}(t,{\bf
x})+\sum_{n=2}^{\infty}{1\over n!}\zeta^{(n)}(t,{\bf x}). \q We
assume that the first-order term $\zeta^{(1)}$ is Gaussian and
higher-order terms describe the non-Gaussianity of the full
nonlinear $\zeta$. Working in the framework of Fourier
transformation of $\zeta$, the primordial power spectrum ${\cal
P}_\zeta$ is defined by \e \langle\zeta({\bf k_1})\zeta({\bf
k_2})\rangle=(2\pi)^3 {\cal P}_{\zeta}(k_1)\delta^3({\bf k_1}+{\bf
k_2}), \q and the primordial bispectrum takes the form \e
\langle\zeta({\bf k_1})\zeta({\bf k_2})\zeta({\bf
k_3})\rangle=(2\pi)^3 B_\zeta({\bf k_1}, {\bf k_2})\delta^3({\bf
k_1}+{\bf k_2}+{\bf k_3}). \q The amplitude of the bispectrum
relative to the power spectrum is parameterized by the
non-Gaussianity parameter $f_{NL}$, i.e. \e B_\zeta({\bf k_1}, {\bf
k_2})={6\over 5} f_{NL}[{\cal P}_\zeta(k_1){\cal P}_\zeta(k_2)+2\
\hbox{perms}]. \label{dfnl}\q

The primordial density perturbation can be described in terms of the
nonlinear curvature perturbation on uniform density hypersurfaces
\cite{Lyth:2004gb} \e \zeta(t,{\bf x})=\delta \cn (t,{\bf x})+
{1\over 3}\int _{{\bar \rho}(t)}^{\rho(t,{\bf x})}{d{\tilde
\rho}\over {\tilde \rho}+{\tilde p}}, \label{dln}\q where $\cn=\int
Hdt$ is the integrated local expansion, ${\bar \rho}$ is the
homogeneous density in the background model, ${\tilde \rho}$ is the
local density and ${\tilde p}$ is the local pressure.

After inflation inflaton decays into radiations which dominate the
total energy density of our universe. In the radiation dominated era
the Hubble parameter $H$ goes like $\sim a^{-2}$. Once the Hubble
parameter drops below the mass of curvaton field the field starts to
oscillate. Nonlinear evolution of the values of curvatons on large
scale is possible if the potential of curvatons deviates from a
purely quadratic potential away from their minimums
\cite{Lyth:2003dt,Enqvist:2005pg}. Thus, in general, the initial
amplitude of curvaton oscillations $\sigma_{i,o}$ is some function
of the field value $\sigma_{i,*}$ at the Hubble exit \footnote{In
this paper the subscript $*$ denotes the quantity evaluated at the
Hubble exit. }: \e \sigma_{i,o}=g_i(\sigma_{i,*}). \q When the
curvaton starts to oscillate about the minimum of its potential, but
before it decays, it behaves like pressureless dust
$(\rho_{\sigma_{i,o}}\sim a^{-3})$ and the nonlinear curvature
perturbation on uniform-curvaton density hypersurfaces is given by
\e \zeta_{\sigma_{i,o}}(t,{\bf x})=\delta\cn(t,{\bf x})+\int _{{\bar
\rho}_{\sigma_{i,o}}(t)}^{\rho_{\sigma_{i,o}}(t,{\bf x})}{d{\tilde
\rho_{\sigma_{i,o}}}\over 3{\tilde \rho}_{\sigma_{i,o}}}. \q The
curvaton density on spatially flat hypersurfaces is \e
\rho_{\sigma_{i,o}}|_{\delta \cn=0}=e^{3\zeta_{\sigma_{i,o}}}{\bar
\rho}_{\sigma_{i,o}}.  \label{rhs}\q The quantum fluctuations in a
weakly coupled field, such as curvaton, at Hubble exit during
inflation are expected to be well described by a Gaussian random
field \cite{Seery:2005gb}. So we have \e \sigma_{i,*}={\bar
\sigma}_{i,*}+\delta \sigma_{i,*}, \q without higher-order nonlinear
terms. During the curvaton oscillation we expand the energy density
$\rho_{\sigma_{i,o}}=\half m_i^2\sigma_{i,o}^2$
and $\zeta_{\sigma_{i,o}}$ to second order: \m \rho_{\sigma_{i,o}}&=&{\bar \rho}_{\sigma_{i,o}} \[1+2X_i+(1+h_i)X_i^2\], \\
\zeta_{\sigma_{i,o}}&=&\zeta_{\sigma_{i,o}}^{(1)}+\half
\zeta_{\sigma_{i,o}}^{(2)}, \n where ${\bar
\rho}_{\sigma_{i,o}}=\half m_i^2 {\bar \sigma}_{i,o}^2$, ${\bar
\sigma}_{i,o}\equiv g_i(\sigma_{i,*})$, and \m X_i&=&{\delta
\sigma_{i,o}^{(1)}\over {\bar \sigma}_{i,o}}, \label{xx} \\
h_i&=&{g_ig_i''\over g_i'^2}. \n Here prime denotes the derivative
with respect to $\sigma_{i,*}$. Order by order, from Eq.(\ref{rhs})
we have
\m \zeta_{\sigma_{i,o}}^{(1)}&=& {2\over 3}X_i, \label{zs} \\
\zeta_{\sigma_{i,o}}^{(2)}&=&-{3\over
2}\(1-h_i\)\(\zeta_{\sigma_{i,o}}^{(1)}\)^2. \label{zss}\n

In N-vaton model, we assume that the curvatons have the same decay
rate $\Gamma_\sigma$. When the Hubble parameter drops below
$\Gamma_\sigma$, all of curvatons decay into radiations. In order to
get analytic expressions, we work in the sudden-decay approximation
which means that all of curvatons suddenly decay into radiations at
the time $t_D$ when $H=\Gamma_\sigma$. For simplicity, we assume
$m_i>\Gamma_\sigma$ for $i=1,2,...,N$ and then all of curvatons
begin oscillating before they decay.

The curvatons-decay hypersurface is a uniform-density hypersurface
and thus from Eq.(\ref{dln}) the perturbed expansion on this
hypersurface is $\delta \cn=\zeta$, where $\zeta$ is the total
curvature perturbation at curvatons-decay hypersurface. Before the
curvatons decay, there have been radiations produced by decay of
inflaton. Since the equation of state of radiation is
$p_r=\rho_r/3$, the curvature perturbation related to radiations is
\e \zeta_r=\zeta+{1\over 4}\ln{\rho_r\over {\bar \rho}_r}. \q The
curvatons behave like pressureless dust $(p_{\sigma_i}=0)$ and thus
\e \zeta_{\sigma_{i,o}}=\zeta+{1\over 3}\ln{\rho_{\sigma_{i,o}}\over
{\bar \rho}_{\sigma_{i,o}}}. \q In the absence of interations, the
curvature perturbations $\zeta_r$ and $\zeta_{\sigma_{i,o}}$ are
conserved respectively and the above two equations can be written as
\m
\rho_r&=&{\bar \rho}_r e^{4(\zeta_r-\zeta)}, \\
\rho_{\sigma_{i,o}}&=&{\bar
\rho}_{\sigma_{i,o}}e^{3(\zeta_{\sigma_{i,o}}-\zeta)}. \n At the
time of curvatons decay, the total energy density $\rho_{tot}$ is
conserved, i.e. \e \rho_r(t_D,{\bf
x})+\sum_{i=1}^N\rho_{\sigma_{i,o}}(t_D,{\bf x})={\bar
\rho}_{tot}(t_D). \q Requiring that the total energy density is
uniform on the decay surface, we have \e
\(1-\Omega_{\sigma,D}\)e^{4(\zeta_r-\zeta)}+
\sum_{i=1}^{N}\Omega_{\sigma_i,D}e^{3(\zeta_{\sigma_{i,o}}-\zeta)}=1,
\label{nlc}\q where $\Omega_{\sigma_i,D}={\bar
\rho}_{\sigma_{i,D}}/{\bar \rho}_{tot}$ is the fraction of curvaton
energy density in the energy budget at the time of curvaton decay,
and \e \Omega_{\sigma,D}\equiv\sum_{i=1}^{N}\Omega_{\sigma_i,D}.\q
Actually $\zeta_r$ is generated by the fluctuation of inflaton
$\phi$ during inflation, namely $\zeta_r=\zeta_\phi$. In N-vaton
scenario, usually we assume the curvature perturbation caused by
inflaton is relatively small and can be neglected. The more general
case with $\zeta_r=\zeta_\phi\neq 0$ is discussed in Appendix A.
Here we consider $\zeta_r=0$ and Eq.(\ref{nlc}) gives \e e^{4\zeta}-
\(\sum_{i=1}^N \Omega_{\sigma_i,D}
e^{3\zeta_{\sigma_{i,o}}}\)e^\zeta+\Omega_{\sigma,D}-1=0.
\label{nl}\q Order by order, from Eq.(\ref{nl}) we have \e
\zeta^{(1)}=A\sum_{i=1}^N
\Omega_{\sigma_i,D}\zeta_{\sigma_{i,o}}^{(1)}, \label{zf}\q and \e
\zeta^{(2)}={1\over 4-\Omega_{\sigma,D}}\[{9\over 2} \sum_{i=1}^N
\Omega_{\sigma_i,D}(1+h_i)
\(\zeta_{\sigma_{i,o}}^{(1)}\)^2-(8+\Omega_{\sigma,D})\(\zeta^{(1)}\)^2\],
\q where \e A={3\over 4-\Omega_{\sigma,D}}. \q The total curvature
perturbation up to second order is \m \zeta&=&\zeta^{(1)}+\half
\zeta^{(2)}=A\sum_{i=1}^N
\Omega_{\sigma_i,D}\zeta_{\sigma_{i,o}}^{(1)} \nonumber \\
&+&{3A\over 4}\sum_{i=1}^N \Omega_{\sigma_i,D}(1+h_i)
\(\zeta_{\sigma_{i,o}}^{(1)}\)^2-(1+{A\over
2}\Omega_{\sigma,D})A^2\(\sum_{i=1}^N\Omega_{\sigma_i,D}\zeta_{\sigma_{i,o}}^{(1)}\)^2.
\label{zz}\n

Assume that the two different curvatons are uncorrelated with each
other and then \e \langle\zeta_{\sigma_{i,o}}^{(1)}({\bf k_1})
\zeta_{\sigma_{j,o}}^{(1)}({\bf k_2})\rangle=(2\pi)^3 {\cal
P}_{\zeta_{\sigma_{i,o}}}(k_1)\delta_{ij}\delta^3({\bf k_1}+{\bf
k_2}). \q Using Eq.(\ref{zf}), we can easily calculate the
primordial power spectrum: \e {\cal P}_\zeta=A^2\sum_{i=1}^N
\Omega_{\sigma_i,D}^2{\cal P}_{\zeta_{\sigma_{i,o}}}. \label{cp} \q
For convenience, we introduce a new parameter $\alpha_i$ as follows
\e {\cal P}_{\zeta_{\sigma_{i,o}}}=A^{-2}\alpha_i{\cal P}_\zeta.
\label{cpz}\q The constraint on the coefficients $\alpha_i$ is \e
\sum_{i=1}^N \Omega_{\sigma_i,D}^2\alpha_i=1. \label{ma}\q Similarly
we can also calculate the primordial bispectrum: \m
\langle\zeta({\bf k_1})\zeta({\bf k_2})\zeta({\bf k_3})\rangle &=&
{3A^3\over 4}\sum_{i,j,k=1}^N
\Omega_{\sigma_i,D}\Omega_{\sigma_j,D}\Omega_{\sigma_k,D} \langle
\zeta_{\sigma_{i,o}}^{(1)}({\bf k_1})\zeta_{\sigma_{j,o}}^{(1)}({\bf
k_2}) (\zeta_{\sigma_{k,o}}^{(1)}*\zeta_{\sigma_{k,o}}^{(1)})({\bf
k_3})\rangle \nonumber \\ &-&(1+{A\over
2}\Omega_{\sigma,D})A^4\sum_{i,j,k,l=1}^N
\Omega_{\sigma_i,D}\Omega_{\sigma_j,D}\Omega_{\sigma_k,D}\Omega_{\sigma_l,D}
\nonumber \\ &\times& \langle \zeta_{\sigma_{i,o}}^{(1)}({\bf
k_1})\zeta_{\sigma_{j,o}}^{(1)}({\bf k_2})
(\zeta_{\sigma_{k,o}}^{(1)}*\zeta_{\sigma_{l,o}}^{(1)})({\bf
k_3})\rangle \nonumber \\ &+& 2\ \hbox{permutations of}\ \{{\bf
k_1}, {\bf k_2}, {\bf k_3} \}, \n where $*$ denotes a convolution as
follows \e
(\zeta_{\sigma_{i,o}}^{(1)}*\zeta_{\sigma_{j,o}}^{(1)})({\bf
k})={1\over (2\pi)^3}\int d^3{\bf q}\zeta_{\sigma_{i,o}}^{(1)}({\bf
q})\zeta_{\sigma_{j,o}}^{(1)}({\bf k}-{\bf q}). \q After
straightforward calculations, we get \m \langle\zeta({\bf
k_1})\zeta({\bf k_2})\zeta({\bf k_3})\rangle
&=& \[{3\over 2A}\sum_{i=1}^N\Omega_{\sigma_i,D}^3\alpha_i^2(1+h_i)-(2+A\Omega_{\sigma,D})\] \nonumber \\
&\times& (2\pi)^3[{\cal P}_\zeta(k_1){\cal P}_\zeta(k_2)+2 \
perms]\delta^3({\bf k_1}+{\bf k_2}+{\bf k_3}). \n Using the
definition of $f_{NL}$ in Eq.(\ref{dfnl}), we have \e f_{NL}={5\over
4A}\sum_{i=1}^N\Omega_{\sigma_i,D}^3\alpha_i^2(1+h_i)-({5\over
3}+{5A\over 6}\Omega_{\sigma,D}). \label{fnlo}\q For single
curvaton, the solution of Eq.(\ref{ma}) is
$\alpha=1/\Omega_{\sigma,D}^2$ and then $f_{NL}^{single}={5\over
4f_D}(1+h)-{5\over 3}-{5f_D\over 6}$, where
$f_D=A\Omega_{\sigma,D}$. It is the same as the result in the
literatures.

To compare with the cosmological observations, we introduce a
``dimensionless" power spectrum $P_{\zeta}$ which is defined by \e
\langle \zeta({\bf k_1})\zeta({\bf k_2})\rangle\equiv {2\pi^2\over
k_1^3}P_\zeta \delta^3({\bf k_1}+{\bf k_2}). \q The power spectrum
of $\delta_{\sigma_{i,*}}$ is given by \e
P_{\delta_{\sigma_{i,*}}}=\({H_*\over 2\pi}\)^2. \q According to
Eq.(\ref{xx}) and (\ref{zs}), we have \e
P_{\zeta_{\sigma_{i,o}}}={4\over 9}q_i^2
P_{\delta_{\sigma_{i,*}}}={1\over 9\pi^2}q_i^2 H_*^2,\label{pzs} \q
where \e q_i={g_i'/g_i}. \q The value of $\alpha_i$ takes the form
\e \alpha_{i}=A^2P_{\zeta_{\sigma_{i,o}}}/P_\zeta={A^2\over
9\pi^2}{q_i^2H_*^2\over P_\zeta}. \label{api}\q Based on
Eq.(\ref{cp}), the amplitude of the primordial power spectrum
$P_\zeta$ becomes \e P_\zeta={A^2\over 9\pi^2}\sum_{i=1}^N
\Omega_{\sigma_i,D}^2 q_i^2 H_*^2. \label{pzo} \q In
\cite{Komatsu:2008hk} WMAP normalization of the primordial power
spectrum is \e P_{\zeta,WMAP}=2.457_{-0.093}^{+0.092}\times 10^{-9}.
\label{wpz}\q On the other hand, the amplitude of primordial power
spectrum generated by inflaton is \e
P_{\zeta_\phi}={H_*^2/M_p^2\over 8\pi^2 \epsilon}, \q which should
be much smaller than $P_{\zeta,WMAP}$, namely \e H_*\ll 4.4\times
10^{-4}\sqrt{\epsilon}M_p. \q

Gravitational wave perturbation (tensor perturbation) is also
generated during inflation in N-vaton model. The tensor perturbation
only depends on the inflation scale and its amplitude is given by \e
P_T={H_*^2/M_p^2\over \pi^2/2}. \label{ts}\q Usually we define a new
parameter named the tensor-scalar ratio $r$ to measure the amplitude
of the tensor perturbations: \e r={P_T/P_\zeta}. \q So the Hubble
parameter during inflation is related to the tensor-scalar ratio by
\e H_*={\pi\over \sqrt{2}} P_\zeta^{1/2} r^\half M_p. \label{hrm}\q
Using WMAP normalization (\ref{wpz}), we get $H_*=10^{-4}r^\half
M_p$. If the density perturbation is dominated by inflaton
fluctuation, we have $r=16\epsilon$. In curvaton/N-vaton scenario,
the density perturbation caused by inflaton is subdominant, and thus
the inflation scale should be relatively low, i.e. $r<16\epsilon$.

\vspace{2mm}

{\bf Note}: Going beyond the sudden-decay approximation, the
single-curvaton model was studied in
\cite{Lyth:2002my,Sasaki:2006kq} in detail. We hope that one can do
it for N-vaton model in the future. In this paper we always adopt
the sudden-decay approximation.

\setcounter{equation}{0}
\section{Bound on the non-Gaussianity parameter $f_{NL}$}

In this section we consider the case in which the Hubble parameter
is roughly a constant during inflation. If $\epsilon=-\dot H/H^2$ is
large, the variation of inflaton is larger than the Planck scale
\cite{Lyth:1996im}. Usually this kind of inflation model cannot be
embedded into string theory
\cite{Huang:2007gk,Huang:2007qz,Huang:2007st,Chen:2006hs,Baumann:2006cd,McAllister:2007bg}.
For simplicity we also assume the values of curvatons don't evolve
between Hubble exit during inflation and the beginning of their
oscillations. So we have
$\sigma_{i,o}=g_i(\sigma_{i,*})=\sigma_{i,*}$, and thus $q_i={1/
\sigma_{i,*}}$ and $h_i=0$. Here we are interested in the case of
large non-Gaussianity. From Eq.(\ref{fnlo}), a large non-Gaussianity
can be obtained if $\alpha_i\gg 1$, but $\Omega_{\sigma,D}$ is not
necessarily required to be much smaller than 1. For example, if one
or more coefficients $\alpha_i$ are large enough and
$\Omega_{\sigma_i,D}^2\alpha_i$ takes a finite value for
$\Omega_{\sigma_i,D}\ll 1$, $f_{NL}$ can be large even when
$\Omega_{\sigma,D}=1$ because of $f_{NL}\sim
(\Omega_{\sigma_i,D}^2\alpha_i)^2/\Omega_{\sigma_i,D}$. This case
can be possibly achieved only for multiple curvatons. In the
single-curvaton model, $\Omega_{\sigma,D}^2\alpha=1$ and $f_{NL}\sim
1/\Omega_{\sigma,D}$. However whether the above conditions can be
naturally realized in a concrete N-vaton model is still an open
question and we will return to this problem in some future work.
Here we only give a brief discussion for the case with two curvatons
in the Appendix B.

From now on, we only focus on the case with $\Omega_{\sigma,D}\ll 1$
for simplicity. In this case a large non-Gaussianity is also
expected. Now $A=3/4$, the amplitude of primordial power spectrum
and the non-Gaussianity parameter in Eq.(\ref{pzo}) and (\ref{fnlo})
are respectively simplified to be \e P_\zeta={1\over
16\pi^2}\sum_{i=1}^N \Omega_{\sigma_i,D}^2 {H_*^2\over
\sigma_{i,*}^2}, \label{pz} \q and \e f_{NL}={5\over
3}\sum_{i=1}^N\Omega_{\sigma_i,D}^3\alpha_i^2. \label{fnl}\q Since
$\alpha_i$ is only constrained by Eq.(\ref{ma}), usually we need
more information if we want to constrain the non-Gaussianity
parameter $f_{NL}$.

\subsection{Lower bound on $f_{NL}$}

Let's introduce a very useful inequality \e \sum_{i=1}^N u_i^2 \cdot
\sum_{j=1}^N v_i^2 \geq \(\sum_{i=1}^N {u_iv_i}\)^2, \label{u}\q
where $u_i\geq 0$ and $v_i \geq 0$ for $i=1,2,...,N$. The equality
in Eq.(\ref{u}) is satisfied only when $u_i/u_j=v_i/v_j$ for
$i,j=1,2,...,N$. Using this inequality, we immediately find \e
\sum_{i=1}^N \Omega_{\sigma_i,D} \sum_{j=1}^N
\Omega_{\sigma_j,D}^3\alpha_j^2\geq \(\sum_{i=1}^N
\Omega_{\sigma_i,D}^2\alpha_i\)^2. \q Taking Eq.(\ref{ma}) into
account, we find the non-Gaussianity parameter $f_{NL}$ in
Eq.(\ref{fnl}) is bounded from below, namely \e f_{NL}\geq {5\over
3\Omega_{\sigma,D}}. \label{fnll} \q The equality is satisfied when
$\alpha_i\Omega_{\sigma_i,D}=\theta$ which is a constant. We can
easily check it. In this special case the solution of Eq.(\ref{ma})
is given by \e \theta=1/\Omega_{\sigma,D}, \q and then \e
\alpha_i={1\over \Omega_{\sigma_i,D}\Omega_{\sigma,D}}. \q
Instituting this solution into Eq.(\ref{fnl}), we get \e
f_{NL}={5\over 3\Omega_{\sigma,D}}. \q Keeping $\Omega_{\sigma,D}$
fixed, the non-Gaussianity parameter $f_{NL}$ in N-vaton model is
not less than that in single-curvaton model.

\subsection{Upper bound on $f_{NL}$}

In this subsection we take more information into account. Because we
only focus on the limit of $\Omega_{\sigma,D}\ll 1$, the radiation
produced by inflaton is always dominant before the curvaton decay.
After that curvatons oscillate around their minimums $\sigma_i=0$
and their energy density decreases as $a^{-3}$. Once the Hubble
parameter drops below $\Gamma_\sigma$, the curvatons energy is
converted into radiations. Similar to the arguments in
\cite{Lyth:2001nq,Lyth:2002my}, the energy density parameter
$\Omega_{\sigma_i,D}$ at the time of curvatons decay is given by \e
\Omega_{\sigma_i,D}={\sigma_{i,*}^2\over 6M_p^2}\({m_i\over
\Gamma_\sigma}\)^\half. \label{ms}\q Instituting the above equation
into Eq.(\ref{pz}), the amplitude of primordial power spectrum
becomes \e P_\zeta={H_*^2\over
(24\pi)^2M_p^4\Gamma_\sigma}\sum_{i=1}^Nm_i\sigma_{i,*}^2,
\label{pl}\q or equivalently, \e \sum_{i=1}^N
m_i\sigma_{i,*}^2=(24\pi)^2P_\zeta {M_p^4\Gamma_\sigma \over H_*^2}.
\label{mss}\q The WMAP normalization gives a constraint on
$\sum_{i=1}^N m_i\sigma_{i,*}^2$.

In this section $A=3/4$, $g_i(\sigma_{i,*})=\sigma_{i,*}$ and
$q_i=1/\sigma_{i,*}$. Eq.(\ref{api}) is simplified to be \e
\alpha_i={1\over 16\pi^2}{H_*^2/\sigma_{i,*}^2\over
P_\zeta},\label{apii} \q and then \e \alpha_i\Omega_{\sigma_{i,D}}=
{r\over 192}\sqrt{m_i\over \Gamma_\sigma}.  \label{apo}\q If $m_i=m$
for $i=1,2,...,N$, $\alpha_i\Omega_{\sigma_{i,D}}$ is a constant and
the inequality in Eq.(\ref{fnll}) is saturated. Now we have \e
\theta={r\over 192}\sqrt{m \over \Gamma_\sigma},\q and \e
f_{NL}={5\over 576}r\sqrt{m\over \Gamma_\sigma}. \label{fnlg} \q In
general different curvatons $\sigma_i$ has different mass $m_i$.
Using Eq.(\ref{fnl}), (\ref{ms}) and (\ref{apii}), we find the
non-Gaussianity parameter $f_{NL}$ takes the form \e f_{NL}=3\times
10^{-7}P_\zeta^{-2}{H_*^4\over M_p^6\Gamma_\sigma^{3/2}}
\sum_{i=1}^N m_i^{3\over 2}\sigma_{i,*}^2. \label{fnlx} \q When
$m_i=m$ for $i=1,2,...,N$, we can easily check that this results is
the same as (\ref{fnlg}).

How to determine the value of $\sigma_{i,*}$ is a crucial problem in
curvaton/N-vaton model. In the literatures $\sigma_{i,*}$ are taken
as free parameters. In classical level it is correct. However for a
scalar field $\chi$ in de Sitter space, if its mass is much smaller
than $H_*$, its Compton wavelength is large compared to the
curvature radius of the background $H_*^{-1}$ and the gravitational
effects may play a crucial role on its behavior. In
\cite{Bunch:1978yq,Vilenkin:1982wt,Linde:1982uu} the authors
explicitly showed that the quantum fluctuation of a light scalar
field $\chi$ with mass $m_\chi$ in de Sitter space gives it a
non-zero expectation value of $\chi^2$ \e
\langle\chi^2\rangle={3H_*^4\over 8\pi^2m_\chi^2}. \label{bdw} \q
This result is reliable for a light scaler field with $m_\chi\ll
\sqrt{2}H_*$ in a long-lived, quasi-de Sitter inflation. Here we
also ignore the possible corrections from the cubic, or higher-power
terms in the curvaton potential. So the typical or average energy
density of the scalar field $\chi$ is ${3H_*^4\over 16\pi^2}$. Since
the masses of curvatons are assumed to be much smaller than $H_*$,
the total energy density of curvatons can be estimated as
${3NH_*^4\over 16\pi^2}$, which implies \e \sum_{i=1}^N
m_i^2\sigma_{i,*}^2={3NH_*^4\over 8\pi^2}. \label{mmss}\q Using the
inequality (\ref{u}), Eq.(\ref{mss}) and (\ref{mmss}), we have \e
\sum_{i=1}^N m_i^{3\over 2}\sigma_{i,*}^2\leq
6\sqrt{6}(NP_\zeta\Gamma_\sigma)^\half H_*M_p^2. \q  We see that the
non-Gaussianity parameter $f_{NL}$ in Eq.(\ref{fnlx}) is bounded
from above: \e f_{NL}\leq 4.41\times 10^{-6} P_\zeta^{-{3\over 2}}
N^\half {H_*^5\over M_p^4\Gamma_\sigma}. \label{fnlu}\q The
inequality (\ref{fnlu}) is saturated when these curvatons fields
have the same mass: $m_1=m_2=...=m_N=m$. Now
$\alpha_i\Omega_{\sigma_i,D}$ is a constant and the lower bound
(\ref{fnll}) is also saturated. One point we want to stress is that
$\Omega_{\sigma,D}$ is not kept fixed. Keeping the inflation scale
$H_*$ (or tensor-scalar ratio $r$), the number of the curvatons $N$
and the curvaton decay rate $\Gamma_\sigma$ fixed, the
non-Gaussianity $f_{NL}$ is maximized in the case where all of the
curvatons have the same mass. We discuss this special case in Sec.
3.4 in detail.

\subsection{Adiabatic condition}

In \cite{Beltran:2008ei} the author pointed out that the curvaton
model is free from the constraint of isocurvature perturbation in
WMAP \cite{Komatsu:2008hk} if the cold dark matter (CDM) is not the
direct decay product of the curvatons and CDM is generated after the
curvatons decay completely. So does N-vaton. Denote $H_{cdm}$ as the
Hubble parameter when CDM is generated and thus
$H_{cdm}<\Gamma_\sigma$. The Hubble parameter $H_{cdm}$ is related
to the temperature $T_{cdm}$ at the epoch of CDM creation by
$H_{cdm}=T_{cdm}^2/M_p$. Therefore \e \Gamma_\sigma>{T_{cdm}^2\over
M_p}. \q Combining with Eq.(\ref{fnlu}), we find \e
T_{cdm}<1.87\times 10^{12} N^{1\over 4}r^{5\over 4}f_{NL}^{-\half} \
\hbox{GeV}, \q where we use Eq.(\ref{hrm}) and WMAP normalization
$P_\zeta=P_{\zeta,WMAP}$. In \cite{Jungman:1995df} the relationship
between $T_{cdm}$ and the mass of CDM $M_{cdm}$ is roughly given by
\e M_{cdm}\simeq 20T_{cdm}. \q So the mass of CDM is bounded from
above \e M_{cdm}< 3.7\times 10^{13} N^{1\over 4}r^{5\over
4}f_{NL}^{-\half} \ \hbox{GeV}.\q For example, $N\sim 10^3$, $r\sim
10^{-4}$ and $f_{NL}\sim 50$, the mass of CDM is less than $3\times
10^8$ GeV. On the other hand, $f_{NL}$ is bounded by $1/M_{cdm}^2$
from above.

\subsection{The case with $m_i=m$ for $i=1,2,...,N$}

In this case the constraint coming from the amplitude of power
spectrum (\ref{mss}) and the estimation of the total energy density
of curvaton during inflation (\ref{mmss}) are respectively
simplified to be \e \sigma_T^2\equiv\sum_{i=1}^N \sigma_{i,*}^2=
(24\pi)^2P_\zeta{M_p^4\Gamma_\sigma \over H_*^2m}, \q and \e
\sigma_T^2={3NH_*^4\over 8\pi^2 m^2}.\q According to the above two
equations, we find that curvaton mass $m$ is related to curvaton
decay rate $\Gamma_\sigma$ by \e m=6.68\times 10^{-6}P_\zeta^{-1}
N{H_*^6\over M_p^4\Gamma_\sigma}.\label{mdc}\q Keeping
$\Gamma_\sigma$ fixed, the mass of curvatons $m$ in N-vaton is $N$
times of that in single-curvaton model.

In Sec. 3.2, we argue that the upper bound on the non-Gaussianity
parameter $f_{NL}$ in Eq.(\ref{fnlu}) is saturated when the
curvatons have the same mass and now we have \e f_{NL}=4.41\times
10^{-6} P_\zeta^{-{3\over 2}} N^\half {H_*^5\over
M_p^4\Gamma_\sigma}. \label{fnla} \q Keeping $\Gamma_\sigma$ and
$H_*$ fixed, larger the number of curvatons, larger the
non-Gaussianity parameter $f_{NL}$. On the other hand, if $N$ is
fixed, smaller $\Gamma_\sigma$, larger $f_{NL}$. However, similar to
the argument in \cite{Lyth:2003dt}, the curvaton decay rate is
larger than the gravitational strength decay rate, i.e. \e
\Gamma_\sigma>{1\over c^4}{m^3\over M_p^2}, \label{gg}\q where $c$
is supposed to be an order one coefficient which we have not known
exactly. The curvaton decay rate cannot be arbitrary small.
Substituting Eq.(\ref{mdc}) into (\ref{gg}), we find that
$\Gamma_\sigma$ is bounded from below by the number of curvatons \e
\Gamma_\sigma>1.3\times 10^{-4}c^{-1}P_\zeta^{-{3\over 4}}N^{3\over
4}{H_*^{9/2}\over M_p^{7/2}}. \q The lower bound on the curvatons
decay rate rises as the number of curvatons increases. Combing with
Eq.(\ref{fnla}), we find $f_{NL}$ is bounded by the tensor-scalar
ratio from above \e f_{NL}<0.034\cdot c\cdot P_\zeta^{-{3\over
4}}N^{-{1\over 4}} {H_*^{1/2}\over M_p^{1/2}}=10^3\cdot c \cdot
\({r\over N}\)^{1\over 4}. \label{fnlz}\q For $N=1$, our result is
the same as that in \cite{Huang:2008ze} where we go beyond
sudden-decay approximation and ignore the coefficient ${3/ 8\pi^2}$
when we estimated the expectation value of square of the curvaton
field. Here we introduce an order one coefficient $c$ to encode the
uncertain coefficient in the calculations. On the other hand, using
Eq.(\ref{mdc}), (\ref{fnla}) and (\ref{fnlz}), we obtain \e f_{NL}<
{c^{2/ 3}\over P_\zeta^{2/ 3}} \({m/N\over 1.3\times 10^3
M_p}\)^{1\over 3}=c^{2\over 3}\({m/N\over 2 \times 10^4 \
\hbox{GeV}}\)^{1\over 3}. \q Requiring the mass of each curvatons is
smaller than the Hubble parameter $H_*$ leads to another bound on
$f_{NL}$, i.e. \e f_{NL}<2.3\times 10^3\cdot c^{2\over 3}\cdot
{r^{1/6}\over N^{1/3}}.\label{fnlv}\q If $r>2\times 10^4/(c^4N)$,
the constraint in Eq.(\ref{fnlv}) is more restricted than that in
(\ref{fnlz}). To summarize, the bound on the non-Gaussianity
parameter $f_{NL}$ is given by \e f_{NL}<\hbox{min}\[10^3\cdot c
\cdot \({r\over N}\)^{1\over 4}, \ \ 2.3\times 10^3\cdot c^{2\over
3}\cdot {r^{1/6}\over N^{1/3}}\]. \label{fnlf} \q We see the the
constraint on $f_{NL}$ in N-vaton model is more stringent than that
in single-curvaton model. The reason is that the energy density of
each curvaton during inflation is roughly $H_*^4$ and thus the total
energy density of curvatons in N-vaton model is much larger than
single curvaton energy density. Larger the number of curvatons,
larger $\Omega_{\sigma,D}$. Since $f_{NL}\sim 1/\Omega_{\sigma,D}$,
the non-Gaussianity is suppressed in N-vaton model by the number of
curvatons $N$. A reasonable estimation of the number of curvatons in
string theory might be $10^3$ and $r\leq 10^{-3}$ if we require the
variation of inflaton be smaller than Planck scale
\cite{Huang:2007qz}. If so, $f_{NL}\leq 32\cdot c$. Usually $f_{NL}$
in N-vaton model should be less than $10^2$. For $f_{NL}>10$,
$r>10^{-8}N$ and $m>10^7N$ GeV. Typically we have $N\sim 10^3$ and
then $m>10^{10}$ GeV, $r>10^{-5}$ which implies $H_*>10^{12}$ GeV.

\subsection{N-vaton vs. single-curvaton model}

According to previous discussions, the non-Gaussianity parameter
$f_{NL}$ in N-vaton model is larger than that in single-curvaton
model if the decay rate of curvaton is kept fixed, and the maximum
value of $f_{NL}$ in N-vaton is obtained when all of curvatons have
the same mass. The maximum value of $f_{NL}$ is $\sqrt{N}$ times of
that in single-curvaton model. However now the curvaton mass in
N-vaton is $N$ times of that in single-curvaton model. The
requirement that the curvaton decay rate be larger than the
gravitational strength decay rate in N-vaton model becomes much more
stringent than that in single-curvaton model. That is why the upper
bound on $f_{NL}$ in Eq.(\ref{fnlz}) is suppressed by a factor
$1/N^{1\over 4}$.

On the other hand, we consider the mass of different curvaton is
quite different from each other. For simplicity, we estimate
$\sigma_{i,*}\sim H_*^2/m_i$. According to Eq.(\ref{pl}) and
(\ref{fnlx}), the contributions to the amplitude of primordial power
spectrum and non-Gaussianity parameter from curvaton $\sigma_i$ are
respectively $P_{\zeta_i}\sim {H_*^6/(M_p^4\Gamma_\sigma m_i)}$ and
$f_{NL,i}\sim H_*^8/(P_\zeta^2M_p^6\Gamma_\sigma^{3\over
2}m_i^\half)$. If the lightest curvaton $\sigma_L$ is much lighter
than other curvatons, the total primordial power spectrum and
non-Gaussianity are roughly contributed by $\sigma_L$. Now N-vaton
model is reduced to single-curvaton model and $f_{NL}\sim
m_L/(P_\zeta^\half H_*)$. Similarly, requiring
$\Gamma_{\sigma}>m_L^3/(c^4M_p^2)$ yields $f_{NL}<10^3\cdot c \cdot
r^{1\over 4}$.

\setcounter{equation}{0}
\section{Spectral index and non-Gaussianity}

The spectral index of the primordial power spectrum generated by
curvatons is defined as \e n_s^{nc}\equiv 1+{d\ln P_\zeta^{nc}\over
d\ln k}=1-2\epsilon+2\eta_{\sigma\sigma}, \q where \e
\eta_{\sigma\sigma}\equiv\sum_{i=1}^N
\Omega_{\sigma_{i,D}}^2\alpha_i {1\over 3H_*^2}{d^2V(\sigma_i)\over
d\sigma_i^2}=\sum_{i=1}^N \Omega_{\sigma_{i,D}}^2\alpha_i
{m_i^2\over 3H_*^2}. \label{ett}\q If the primordial power spectrum
is dominated by the curvature perturbation generated by curvatons,
$n_s=n_s^{nc}$. The masses of curvatons are assumed to be much
smaller than $H_*$ and then $n_s\simeq 1-2\epsilon$. For $n_s=0.96$,
$\epsilon=0.02$ which might be realized in landscape inflation
\cite{Huang:2007ek,Huang:2008jr,Chialva:2008zw} or the monodromies
\cite{Silverstein:2008sg}. However in this case the Hubble parameter
$H$ during inflation cannot be taken as a constant any more. In
\cite{Huang:2008qf}, we showed that the values of the curvatons
depend on the initial condition of inflation which should be
fine-tuned to achieve the suitable amplitude of primordial power
spectrum and non-Gaussianity parameter $f_{NL}$. It is quite
unnatural. Usually a closely scale-invariant curvature perturbation
generated by curvaton is expected in curvaton/N-vaton scenario.

On the other hand, the spectral index of the primordial power
spectrum generated by inflaton is \e n_s^{inf}\equiv 1+{d\ln
P_\zeta^{inf}\over d\ln k}=1-6\epsilon+2\eta. \q In some inflation
models, $\epsilon\simeq 0$, but the order of magnitude of $\eta$ can
be $-{\cal O}(10^{-1})$ to $-{\cal O}(10^{-2})$. If the inflaton
fluctuation makes a significant contribution to the total primordial
power spectrum, a red-tilted primordial power spectrum is possibly
obtained. We calculate the curvature perturbation for this scenario
in Appendix A. Introduce a parameter $\beta$ to measure the relative
amplitude of power spectrum generated by curvatons: \e
\beta=P_\zeta^{nc}/P_\zeta^{tot}, \q and then \e
P_\zeta^{inf}=(1-\beta)P_\zeta^{tot}. \q Now the spectral index
becomes \m n_s&\equiv&1+{d\ln P_\zeta^{tot}\over d\ln
k}=\beta n_s^{nc}+(1-\beta) n_s^{inf}\nonumber \\
&=&1-(6-4\beta)\epsilon+2\beta\eta_{\sigma\sigma}+2(1-\beta)\eta,
\label{spd}\n where $\alpha_i$ in Eq.(\ref{ett}) should be replaced
by $\gamma_i$. We consider $\epsilon\ll 1$ and
$\eta_{\sigma\sigma}\ll 1$ and thus \e n_s\simeq 1+2(1-\beta)\eta.
\q For $\beta=0.8$ and $\eta=-0.1$, $n_s\simeq 0.96$.

Since the total primordial power spectrum is not only generated by
curvatons in this scenario, some formulations in Sec. 3 should be
modified by some powers of $\beta$: $f_{NL}$ is replaced by
$f_{NL}/\beta^2$ and WMAP normalization becomes $P_\zeta=\beta
P_{\zeta,WMAP}$. For example, Eq.(\ref{fnlf}) is changed to be \e
f_{NL}<\hbox{min}\[10^3\cdot \beta^{5\over 4}\cdot c \cdot \({r\over
N}\)^{1\over 4}, \ \ 2.3\times 10^3\cdot \beta^{4\over 3}\cdot
c^{2\over 3}\cdot {r^{1/6}\over N^{1/3}}\]. \q For $\beta=0.8$, the
bound on the non-Gaussianity does not change so much and a large
value of $f_{NL}$ is still achieved naturally. But we need to stress
that a large value of $f_{NL}$ is obtained only when $r$ is not too
small.

The size of non-Gaussianity generated by inflaton is controlled by a
factor $(1-\beta)^2$ in Eq.(\ref{fnlt}) and rich phenomena are
expected in this mixed scenario. A large value of $f_{NL}^{inf}$ is
possibly detectable if $(1-\beta)$ is not so small. Another bonus of
this mixed scenario is that if the adiabatic fluctuation generated
by inflaton is big compared to that from the curvaton $(\beta\sim
0)$, our model is relaxed from the constraint on the isocurvature
perturbation in \cite{Komatsu:2008hk} even in the case where dark
matter was generated before the decay of the curvaton.

\setcounter{equation}{0}
\section{Random matrix and typical mass spectrum in string theory}

Axions are typically present in large numbers in string
compactifications, and even when all other moduli are stabilized,
the axion potentials remain rather flat as a consequence of
well-known nonrenormalization theorems \cite{Dine:1986vd}. Following
\cite{Dimopoulos:2005ac} the potential of $N$ axions $\varphi_i$ is
\e V(\varphi)=\sum_{i=1}^N \Lambda_i^4\[1-\cos \(\varphi_i\over
f_i\)\], \q where $f_i$ is the axion decay constant and $\Lambda_i$
is the dynamically generated scale of the axion potential that
typically arises from an instanton expansion. Redefine the axion
field as $\sigma_i\equiv\varphi/f_i$ and then the Lagrangian for
small axion displacements $\sigma_i\ll M_p$ in \cite{Easther:2005zr}
is given by \e {\cal L}=\sum_{i=1}^N\[\half (\p \sigma_i)^2-\half
m_i^2\sigma_i^2\]. \q In \cite{Dimopoulos:2005ac,Easther:2005zr} the
axion fields are taken as inflatons and the value of $\sigma_i$ is
larger than $M_p/\sqrt{N}$, but smaller than $M_p$. In
\cite{Huang:2007st} we argued that the vacuum expectation value of
$\sigma_i$ is bounded by $M_p/\sqrt{N}$ from above and this
inflation model might be inconsistent with full quantum theory of
gravity. In this section we suggest that these axion fields play the
role as curvatons, not inflatons.

It is still difficult to explicitly calculate the mass of axion in
the context of KKLT moduli stabilization \cite{Kachru:2003aw}.
However, in \cite{Easther:2005zr} the authors found an essentially
universal probability distribution for the mass square of axions as
the Marcenko-Pastur law \e p(m^2)={1\over 2\pi v}{\sqrt{(b-m^2/{\bar
m}^2)(m^2/{\bar m}^2-a)}\over m^2}, \q for
$a\leq m^2/{\bar m}^2\leq b$, where \m a&=&(1-\sqrt{v})^2, \\
b&=&(1+\sqrt{v})^2. \n The shape of the distribution only depends on
a single parameter $v$ which is determined by the dimensions of the
Kahler and complex structure moduli spaces. This distribution is
universal because it does not depend on specific details of the
compactification, such as the intersection numbers, the choice of
fluxes, or the location in moduli space. It is also insensitive to
superpotential corrections. But we cannot determine the overall mass
scale from string theory. In a KKLT compactification of Type IIB
string theory, there are $h_{1,1}$ axions, and $h_{1,1}+h_{2,1}+1$
is the total dimension of the moduli space (Kahler, complex
structure, and dilaton), so that \e v={h_{1,1}\over
h_{1,1}+h_{2,1}+1}. \q In \cite{Easther:2005zr} the authors argued
that the models of $v=\half$ are strongly favored.

In general, the number of axions is roughly ${\cal O}(10^2\sim
10^3)$ in string theory. So \e {1\over N}\sum_{i=1}^N m_i^{2k}\equiv
\langle m^{2k}\rangle \simeq \int_{a{\bar m}^2}^{b{\bar
m}^2}m^{2k}p(m^2)dm^2={\bar m}^{2k}s(k,v) \q up to the order of
$1/N$ which can be safely neglected in our analysis, where $k$ is
just a number and \e s(k,v)={1\over 2\pi
v}\int_a^bx^{k-1}\sqrt{(b-x)(x-a)}dx. \q The function $s(k,v)$ has
some interesting properties: \e s(0,v)=s(1,v)=s(k,0)=1. \q Since
$s(1,v)=1$, $\langle m^2\rangle={\bar m}^2$ which denotes the
overall mass scale. We also define \e {1\over N}\sum_{i=1}^N
m_i^{2k}\sigma_i^2 \equiv \langle m^{2k}\sigma^2\rangle=\langle
m^{2k}\langle\sigma^2\rangle\rangle. \q Here we have
$\langle\sigma^2\rangle= {3H_*^4\over 8\pi^2 m^2}$ and then \e
\sum_{i=1}^N m_i^{2k}\sigma_i^2={3N\over 8\pi^2}H_*^4{\bar
m}^{2(k-1)}s(k-1,v). \q Because $s(0,v)=1$, $\sum_{i=1}^N
m_i^2\sigma_i^2= {3NH_*^4\over 8\pi^2}$ and Eq.(\ref{mmss}) is
automatically satisfied. Here we assume that $\sigma_i\ll f_i$.
Otherwise, the quartic, or higher-power correction terms from the
expansion of the axion potential will be important. In string
theory, the axion decay constant $f_i$ is generically large
\cite{Svrcek:2006yi} and our assumption of $\sigma_i\ll f_i$ is
reasonable.

According to Eq.(\ref{pl}) and (\ref{fnlx}), we can easily calculate
the amplitude of primordial power spectrum and the non-Gaussianity
parameter generated by curvatons: \m P_\zeta^{nc}&=& 6.68\times 10^{-6}s(-1/2,v){NH_*^6\over M_p^4\Gamma_\sigma {\bar m}}, \\
f_{NL}^{nc}&=& 1.14\times
10^{-8}s(-1/4,v)(P_\zeta^{nc})^{-2}{NH_*^8\over
M_p^6\Gamma_\sigma^{3\over 2}{\bar m}^\half}. \n Here are three
unknown scales, $H_*$, $\Gamma_\sigma$ and ${\bar m}$, which still
cannot be determined by microscopic physics. Canceling
$\Gamma_\sigma$, we have \e
f_{NL}^{nc}=0.66f_1(v)(P_\zeta^{nc})^{-\half}N^{-\half}{{\bar
m}\over H_*}, \label{nfp}\q where \e f_1(v)=s(-1/4,v)/s^{3\over
2}(-1/2,v). \q On the other hand, canceling ${\bar m}$ yields \e
f_{NL}^{nc}=4.41\times 10^{-6}f_2(v)(P_\zeta^{nc})^{-{3\over
2}}N^\half {H_*^5\over M_p^4\Gamma_\sigma}, \q with \e
f_2(v)=s(-1/4,v)/s^{1\over 2}(-1/2,v). \q We have $f_1(1/2)=0.76$
and $f_2(1/2)=0.98$. The behaviors of $f_1(v)$ and $f_2(v)$ are
showed in Fig. 1.
\begin{figure}[h]
\begin{center}
\leavevmode \epsfxsize=0.6\columnwidth \epsfbox{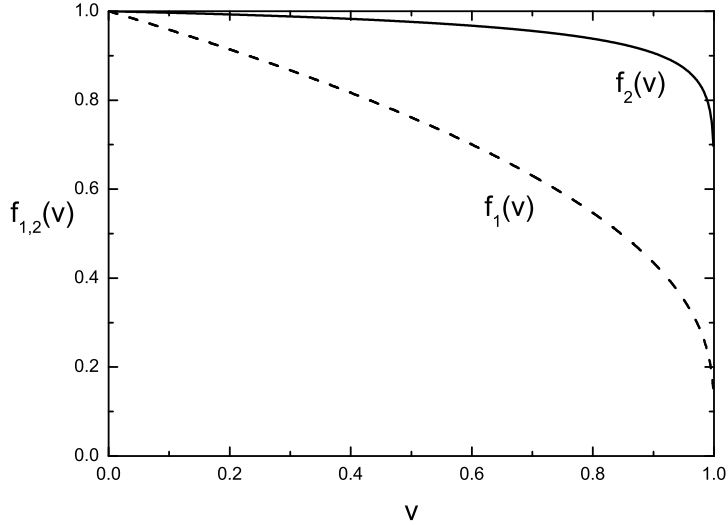}
\end{center}
\caption{The function $f_1(v)$ and $f_2(v)$.}
\end{figure}
When $v\rightarrow 0$, the mass gap of curvatons disappears and we
can expect that our model is reduced to to the case in Sec. 3.4
where all of curvatons have the same mass. Since $s(k,0)=1$ and then
$f_1(0)=f_2(0)=1$, we see both the amplitude of primordial power
spectrum and the non-Gaussianity parameter are really the same as
those in Sec. 3.4. On the other hand, in Sec. 3.2, we find that the
non-Gaussianity parameter $f_{NL}$ is maximized when
$m_1=m_2=...=m_N=m$ for keeping $H_*$, $\Gamma_\sigma$ and $N$
fixed. This model is really consistent with our analysis: $f_2(v)$
approaches its maximum value when $v\rightarrow 0$. Here we also
know how the mass scale ${\bar m}$ varies with $v$. For a given
$f_{NL}^{nc}$, ${\bar m}\sim 1/f_1(v)$ rises as $v$ increases.

In this case, we can also calculate $\eta_{\sigma\sigma}$, i.e. \e
\eta_{\sigma\sigma}={1\over 3} \tau(v){{\bar m}^2\over H_*^2},
\label{et} \q where \e \tau(v)=s(1/2,v)/s(-1/2,v). \q The function
$\tau(v)$ is illustrated in Fig. 2 and $\tau(1/2)=0.73$.
\begin{figure}[h]
\begin{center}
\leavevmode \epsfxsize=0.6\columnwidth \epsfbox{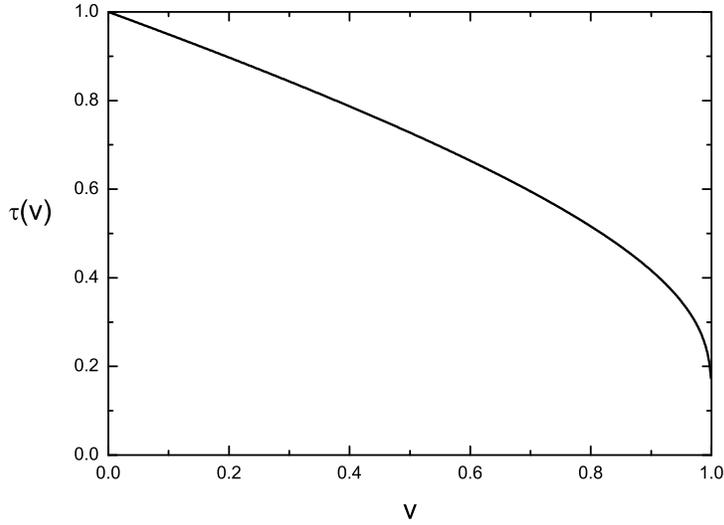}
\end{center}
\caption{The function $\tau(v)$.}
\end{figure}

In general, curvatons fluctuations only contribute to a part of the
total primordial power spectrum, $P_\zeta^{nc}=\beta P_\zeta^{tot}$,
and then $f_{NL}\simeq \beta^2 f_{NL}^{nc}$. Similarly, requiring
$\Gamma_\sigma>{\bar m}^3/(c^4M_p^2)$ yields \e
f_{NL}<\hbox{min}\[10^3\cdot \beta^{5\over 4}\cdot c \cdot
d_1(v)\cdot \({r\over N}\)^{1\over 4}, \ \ 2.3\times 10^3\cdot
\beta^{4\over 3}\cdot c^{2\over 3}\cdot
d_2(v)\cdot {r^{1/6}\over N^{1/3}}\], \q where \m d_1(v)&=& s(-1/4,v)/s^{5\over 4}(-1/2,v), \\
d_2(v)&=& s(-1/4,v)/s^{4\over 3}(-1/2,v). \n The functions $d_1(v)$
and $d_2(v)$ are shown in Fig. 3, and $d_1(1/2)=0.81$ and
$d_2(v)=0.79$.
\begin{figure}[h]
\begin{center}
\leavevmode \epsfxsize=0.6\columnwidth \epsfbox{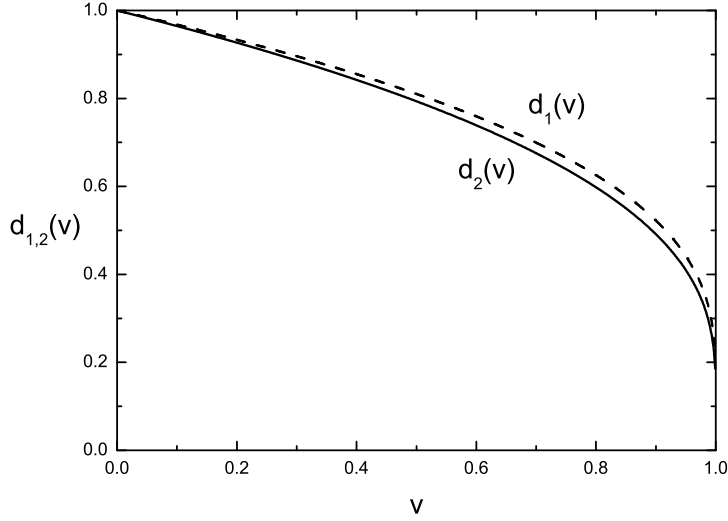}
\end{center}
\caption{The function $d_1(v)$ and $d_2(v)$.}
\end{figure}
Again we see that our results are exactly reduced to the case in
Sec. 3.4 when the parameter $v$ approaches to zero.

\subsection{Compare to experiments}

In this subsection we focus on how to compare our model to
experiments. We consider the case with $P_\zeta^{nc}=\beta
P_\zeta^{tot}$. There are 8 parameters: \m \hbox{inflation}&:& \quad
H_*, \epsilon, \eta \nonumber \\
\hbox{N-vaton}&:& \quad N, \Gamma_\sigma, {\bar m}, v \nonumber \\
\hbox{ratio parameter}&:& \quad \beta \nonumber \n Using
Eq.(\ref{et}), we have ${{\bar m}\over
H_*}=\sqrt{3\eta_{\sigma\sigma}/\tau(v)}$ and then \e
f_{NL}=1.14\beta^{3\over
2}f_1(v)(P_\zeta^{tot})^{-\half}N^{-\half}\sqrt{\eta_{\sigma\sigma}/\tau(v)}.
\q Since $P_\zeta^{inf}={H_*^2/M_p^2\over 8\pi
\epsilon}=(1-\beta)P_\zeta^{tot}$, the relationship between
tensor-scalar ration $r$ and $\epsilon$ is given by \e
r=16(1-\beta)\epsilon. \q The spectral index is given in
Eq.(\ref{spd}) as \e
n_s=1-(6-4\beta)\epsilon+2\beta\eta_{\sigma\sigma}+2(1-\beta)\eta.
\q

To summarize, there are four quantities which can be measured by
experiments: $P_\zeta^{tot}$,
$f_{NL}(\beta,v,N,\eta_{\sigma\sigma})$, $r(\beta,\epsilon)$ and
$n_s(\beta,\epsilon,\eta,\eta_{\sigma\sigma})$. For a given
inflation model, which means $\epsilon$ and $\eta$ are given, the
parameters $\eta_{\sigma\sigma}$, $N$ and $\beta$ can be determined
by experiments for the preferable model with $v=1/2$. Furthermore,
if the number of curvatons is given by the string theory, we can
check whether our model is consistent with experiments. For example,
Let's consider an inflation model with $\epsilon=10^{-4}$ and
$\eta=-0.1$, and we also assume ${\bar m}/H_*\ll 1$ (or
$\eta_{\sigma\sigma}\ll 1$). The tensor perturbation is too small to
be detected. For $n_s=0.96$, $P_\zeta^{tot}=2.457\times 10^{-9}$,
$f_{NL}=30$, $v=1/2$ and $N=10^3$, we find $\beta=0.8$ and
$\eta_{\sigma\sigma}=0.0042$. Now $r=3.2\times 10^{-4}$,
$H_*=4.36\times 10^{12}$ GeV and then ${\bar m}=5.7\times 10^{11}$
GeV.

\setcounter{equation}{0}
\section{Discussions}

In this paper we explicitly calculate the primordial curvature
perturbation in N-vaton model. Multiplicity of light scalar fields
is generic in the theories going beyond standard model. Even though
the total energy density of these light scalar fields is subdominant
during inflation, the perturbation produced by them can dominate the
density perturbation on large scale. We also suggest a realistic
N-vaton model in which the axions in the KKLT compactification of
Type IIB string theory are taken as curvatons, and a rich
phenomenology is shown up. If a large local-type non-Gaussianity is
confirmed by the forthcoming experiments, it can shed a light on
these light scalar fields.

In order to fit the spectral index from WMAP data, the inflaton
fluctuation is still required to play a significant role in the
total primordial power spectrum. Generally the tensor-scalar ratio
is required to be not smaller than $10^{-5}$ if $f_{NL}^{local}>10$.
Many inflation models constructed in string theory, such as brane
inflation \cite{Dvali:1998pa,Kachru:2003sx}, happen in a quite low
energy scale with $r \sim 10^{-10}$ which is too small to generate a
large non-Gaussianity in curvaton/N-vaton scenario. How to construct
an inflation model with $r\sim {\cal O}(10^{-5}-10^{-3})$ and
$\eta\sim -0.1$ is still an open question.

In general, the curvaton decay rate mediated by particles of mass
$M_X$ is expected to be of order $m^3/M_X^2$. So it is natural to
assume that the different curvatons have different decay rates and
decay at different time, in particular for the case where they have
different masses. The authors in \cite{Assadullahi:2007uw} gave a
concrete example to show that a large non-Gaussianity can be
obtained even when the curvatons dominate the total energy density
at the time of decays in the case of two curvatons. However if there
are hundreds or thousands of curvatons, whether this enhancement of
the non-Gaussianity is generic or not is still an open question. It
is worth studying this problem in the future.

\vspace{.5cm}

\noindent {\bf Acknowledgments}

We would like to thank Daniel Chung, Hironobu Kihara, Han-Tao Lu,
K.~P.~Yogendran and Yu-Feng Zhou for useful discussions.

\newpage

\appendix

\setcounter{equation}{0}
\section{Appendix: $\zeta_r=\zeta_\phi\neq 0$}

This is the most general case. We also expand the curvature
perturbation $\zeta_\phi$ to second order as follows \e
\zeta_{\phi}=\zeta_{\phi}^{(1)}+\half \zeta_{\phi}^{(2)}.\q Now
Eq.(\ref{nlc}) becomes \e e^{4\zeta}=\(\sum_{i=1}^N
\Omega_{\sigma_i}
e^{3\zeta_{\sigma_{i,o}}}\)e^\zeta+(1-\Omega_{\sigma,D})e^{4\zeta_\phi}.
\q Order by order, from the above equation we have \e
\zeta^{(1)}=A\sum_{i=1}^N
\Omega_{\sigma_i,D}\zeta_{\sigma_{i,o}}^{(1)}+B\zeta_\phi^{(1)},\q
and \m \zeta^{(2)}&=&{3A\over 2} \sum_{i=1}^N
\Omega_{\sigma_i,D}(1+h_i)
\(\zeta_{\sigma_{i,o}}^{(1)}\)^2-(2+A\Omega_{\sigma,D})A^2\(\sum_{i=1}^N\Omega_{\sigma_{i,D}}\zeta_{\sigma_{i,o}}^{(1)}\)^2
\nonumber \\ &-&8A^2B\sum_{i=1}^N
\Omega_{\sigma_i,D}\zeta_{\sigma_{i,o}}^{(1)}\zeta_\phi^{(1)}+B^2C\(\zeta_\phi^{(1)}\)^2+B\zeta_\phi^{(2)},
\n where \e A={3\over 4-\Omega_{\sigma,D}}, \quad
B={1-\Omega_{\sigma,D}\over 1-\Omega_{\sigma,D}/4}, \quad \hbox{and}
\quad C={3\Omega_{\sigma,D}\over 1-\Omega_{\sigma,D}}A. \q The total
curvature perturbation is given by \e \zeta=\zeta^{(1)}+\half
\zeta^{(2)}. \q

We assume all of these fields including curvatons and inflaton are
independent. Thus the two different curvatons are uncorrelated with
each other \e \langle\zeta_{\sigma_{i,o}}^{(1)}({\bf k_1})
\zeta_{\sigma_{j,o}}^{(1)}({\bf k_2})\rangle=(2\pi)^3 {\cal
P}_{\zeta_{\sigma_{i,o}}}(k_1)\delta_{ij}\delta^3({\bf k_1}+{\bf
k_2}), \q and curvatons are also decoupled to inflaton $\phi$ \e
\langle\zeta_{\sigma_{i,o}}^{(1)}({\bf k_1}) \zeta_{\phi}^{(1)}({\bf
k_2})\rangle=0. \q The primordial power spectrum ${\cal
P}_{\zeta_\phi}$ generated by inflaton is defined as  \e
\langle\zeta_\phi^{(1)}({\bf k_1}) \zeta_\phi^{(1)}({\bf
k_2})\rangle=(2\pi)^3{\cal P}_{\zeta_\phi}\delta^3({\bf k_1}+{\bf
k_2}). \q The fluctuations of curvatons and inflaton contribute to
the total primordial power spectrum which is given by \e {\cal
P}_\zeta^{tot}={\cal P}_\zeta^{nc}+{\cal P}_\zeta^{inf}, \q where \e
{\cal P}_\zeta^{nc}=A^2\sum_{i=1}^N\Omega_{\sigma_{i,D}}^2{\cal
P}_{\zeta_{\sigma_{i,o}}}, \q is the total curvature perturbation
generated by curvatons and \e {\cal P}_\zeta^{inf}=B^2{\cal
P}_{\zeta_\phi} \q is the curvature perturbation generated by
inflaton. For convenience, we introduce parameters $\beta$ and
$\gamma_i$ as follows \m \beta&=&{\cal P}_\zeta^{nc}/{\cal P}_{\zeta}^{tot}, \\
{\cal P}_{\zeta_{\sigma_{i,o}}}&=&A^{-2}\gamma_i{\cal P}_\zeta^{nc}.
\n Thus we have \e \sum_{i=1}^N\Omega_{\sigma_{i,D}}^2\gamma_i=1. \q
When $\beta=1$, all of the primordial power spectrum is generated by
curvatons and $\gamma_i=\alpha_i$. Now we have \e {\cal
P}_{\zeta_{\sigma_{i,o}}}=A^{-2}\beta\gamma_i {\cal P}_\zeta^{tot},
\q and \e {\cal P}_{\zeta_\phi}=B^{-2}(1-\beta){\cal P}_\zeta^{tot}.
\q Similarly, we can also calculate the total non-Gaussianity
parameter $f_{NL}^{tot}$. Here we only give the result:

\e f_{NL}^{tot}=\beta^2{\tilde f}_{NL}^{nc}+\beta(1-\beta){\tilde
f}_{NL}^{cross}+(1-\beta)^2{\tilde f}_{NL}^{inf}, \label{fnlt}\q
where \m {\tilde f}_{NL}^{nc}&=&{5\over
4A}\sum_{i=1}^N\Omega_{\sigma_i,D}^3\gamma_i^2(1+h_i)-({5\over
3}+{5A\over 6}\Omega_{\sigma,D}), \\ {\tilde
f}_{NL}^{cross}&=&-{20\over 3}A,
\\ {\tilde f}_{NL}^{inf}&=&{5\over 6}C+{1\over B}f_{NL}^{inf}, \n and $f_{NL}^{inf}$ is determined by
concrete inflation models
\cite{ArkaniHamed:2003uz,Chen:2006nt,Bean:2008ga,Li:2008qc,Chen:2007gd,Matsuda:2008hx,Matsuda:2008fk,Gao:2008dt,Li:2008qv,Xue:2008mk}.

\setcounter{equation}{0}
\section{Appendix: Another way to get a large non-Gaussianity in N-vaton model}

In this section, we consider that there are two curvatons whose
masses are $m_1$ and $m_2$ respectively. Without loss of the
generality, we assume $m_1\geq m_2$. Once the Hubble parameter drops
below $m_1$, the curvaton $\sigma_1$ starts to oscillate and its
energy density goes like $\sim a^{-3}$. Similarly, when $H\sim m_2$,
the curvaton $\sigma_2$ begins to oscillate. For simplicity, we
assume that the universe is dominated by radiation before $\sigma_2$
starts to oscillate. The scale factor at the time when $\sigma_1$
starts to oscillate is denoted as $a=1$, and then $a=\sqrt{m_1/
m_2}$ when $H\sim m_2$. Since the energy density of an oscillating
curvaton goes like $\sim a^{-3}$, the ratio of the energy density
between these two curvatons at the time of their decay is \e x\equiv
{\rho_{\sigma_1,D}\over \rho_{\sigma_2,D}}\simeq \sqrt{m_1\over
m_2}{\sigma_{1,*}^2\over \sigma_{2,*}^2}, \q and then \e
\Omega_{\sigma_1,D}={x\over 1+x}\Omega_{\sigma,D}, \quad
\Omega_{\sigma_2,D}={1\over 1+x}\Omega_{\sigma,D}. \q Now we have $
\alpha_i=\alpha_c{H_*^2/\sigma_{i,*}^2}$, where
$\alpha_c=A^2/(9\pi^2P_\zeta)$ is just a numerical coefficient. The
constraint on $\alpha_i$ in Eq.(\ref{ma}) reads \e {1\over
(1+x)^2}\alpha_c\Omega_{\sigma,D}^2{H_*^2\over
\sigma_{2,*}^2}=\(1+\sqrt{m_1\over m_2}x\)^{-1}.\q According to
Eq.(\ref{fnlo}), the non-Gaussianity parameter $f_{NL}$ becomes \e
f_{NL}\simeq {5\over 4A}{1\over \Omega_{\sigma,D}}{(1+x)(1+{m_1\over
m_2}x)\over (1+\sqrt{m_1\over m_2}x)^2}. \q If $m_1=m_2$, $f_{NL}$
is large only when $\Omega_{\sigma,D}\ll 1$. On the other hand, if
$m_1\gg m_2$ and $x\geq {\cal O}(1)$, $f_{NL}\sim
1/(x\Omega_{\sigma,D})$ which is large when $\Omega_{\sigma,D}\ll
1$. Now if $x\ll 1$, but $\omega=\sqrt{m_1/m_2}x\geq {\cal O}(1)$,
we have \e f_{NL}\sim {5\over 4A}{1\over
\Omega_{\sigma,D}}{\omega\over (1+\omega)^2}\sqrt{m_1\over m_2} \q
which can be large even when $\Omega_{\sigma,D}\simeq 1$. However
$x\ll 1$ and $\omega\geq {\cal O}(1)$ cannot be achieved if the
values of curvatons during inflation take the typical values
$\sigma_{i,*}^2={3H_*^4/ (8\pi^2m_i^2)}$, because
$x=(m_2/m_1)^{3/2}$ and thus $\omega=x^{2/3}\ll 1$ if $x\ll 1$. Here
we also want to remind that the assumption that these two curvatons
have the same decay rate is not reasonable if $m_1/m_2 \gg 1$.



\newpage

\begin{thebibliography}{99}
\baselineskip=16pt



\bibitem{Bartolo:2004if}
  N.~Bartolo, E.~Komatsu, S.~Matarrese and A.~Riotto,
  ``Non-Gaussianity from inflation: Theory and observations,''
  Phys.\ Rept.\  {\bf 402}, 103 (2004)
  [arXiv:astro-ph/0406398].


\bibitem{Maldacena:2002vr}
  J.~M.~Maldacena,
  ``Non-Gaussian features of primordial fluctuations in single field
  inflationary models,''
  JHEP {\bf 0305}, 013 (2003)
  [arXiv:astro-ph/0210603].


\bibitem{Komatsu:2008hk}
  E.~Komatsu {\it et al.}  [WMAP Collaboration],
  ``Five-Year Wilkinson Microwave Anisotropy Probe (WMAP)
  Observations:Cosmological Interpretation,''
  arXiv:0803.0547 [astro-ph].


\bibitem{Yadav:2007yy}
  A.~P.~S.~Yadav and B.~D.~Wandelt,
  ``Detection of primordial non-Gaussianity (fNL) in the WMAP 3-year data at
  above 99.5$\%$ confidence,''
  arXiv:0712.1148 [astro-ph].

\bibitem{Komatsu:2001rj}
  E.~Komatsu and D.~N.~Spergel,
  ``Acoustic signatures in the primary microwave background bispectrum,''
  Phys.\ Rev.\  D {\bf 63}, 063002 (2001)
  [arXiv:astro-ph/0005036].


\bibitem{Linde:1996gt}
  A.~D.~Linde and V.~F.~Mukhanov,
  ``Nongaussian isocurvature perturbations from inflation,''
  Phys.\ Rev.\  D {\bf 56}, 535 (1997)
  [arXiv:astro-ph/9610219].

\bibitem{Enqvist:2001zp}
  K.~Enqvist and M.~S.~Sloth,
  ``Adiabatic CMB perturbations in pre big bang string cosmology,''
  Nucl.\ Phys.\  B {\bf 626}, 395 (2002)
  [arXiv:hep-ph/0109214].

\bibitem{Lyth:2001nq}
  D.~H.~Lyth and D.~Wands,
  ``Generating the curvature perturbation without an inflaton,''
  Phys.\ Lett.\  B {\bf 524}, 5 (2002)
  [arXiv:hep-ph/0110002].

\bibitem{Moroi:2001ct}
  T.~Moroi and T.~Takahashi,
  ``Effects of cosmological moduli fields on cosmic microwave background,''
  Phys.\ Lett.\  B {\bf 522}, 215 (2001)
  [Erratum-ibid.\  B {\bf 539}, 303 (2002)]
  [arXiv:hep-ph/0110096].

\bibitem{Lyth:2002my}
  D.~H.~Lyth, C.~Ungarelli and D.~Wands,
  ``The primordial density perturbation in the curvaton scenario,''
  Phys.\ Rev.\  D {\bf 67}, 023503 (2003)
  [arXiv:astro-ph/0208055].

\bibitem{Bartolo:2003jx}
  N.~Bartolo, S.~Matarrese and A.~Riotto,
  ``On non-Gaussianity in the curvaton scenario,''
  Phys.\ Rev.\  D {\bf 69}, 043503 (2004)
  [arXiv:hep-ph/0309033].


\bibitem{Malik:2006pm}
  K.~A.~Malik and D.~H.~Lyth,
  ``A numerical study of non-gaussianity in the curvaton scenario,''
  JCAP {\bf 0609}, 008 (2006)
  [arXiv:astro-ph/0604387].


\bibitem{Sasaki:2006kq}
  M.~Sasaki, J.~Valiviita and D.~Wands,
  ``Non-gaussianity of the primordial perturbation in the curvaton model,''
  Phys.\ Rev.\  D {\bf 74}, 103003 (2006)
  [arXiv:astro-ph/0607627].







\bibitem{Choi:2007fya}
  K.~Y.~Choi and J.~O.~Gong,
  ``Multiple scalar particle decay and perturbation generation,''
  JCAP {\bf 0706}, 007 (2007)
  [arXiv:0704.2939 [astro-ph]].

\bibitem{Assadullahi:2007uw}
  H.~Assadullahi, J.~Valiviita and D.~Wands,
  ``Primordial non-Gaussianity from two curvaton decays,''
  Phys.\ Rev.\  D {\bf 76}, 103003 (2007)
  [arXiv:0708.0223 [hep-ph]].

\bibitem{Valiviita:2008zb}
  J.~Valiviita, H.~Assadullahi and D.~Wands,
  ``Primordial non-gaussianity from multiple curvaton decay,''
  arXiv:0806.0623 [astro-ph].


\bibitem{Huang:2008ze}
  Q.~G.~Huang,
  ``Large Non-Gaussianity Implication for Curvaton Scenario,''
  arXiv:0801.0467 [hep-th].

\bibitem{Ichikawa:2008iq}
  K.~Ichikawa, T.~Suyama, T.~Takahashi and M.~Yamaguchi,
  ``Non-Gaussianity, Spectral Index and Tensor Modes in Mixed Inflaton and
  Curvaton Models,''
  arXiv:0802.4138 [astro-ph].

\bibitem{Multamaki:2008yv}
  T.~Multamaki, J.~Sainio and I.~Vilja,
  ``Non-Gaussianity in three fluid curvaton model,''
  arXiv:0803.2637 [astro-ph].

\bibitem{Suyama:2008nt}
  T.~Suyama and F.~Takahashi,
  ``Non-Gaussianity from Symmetry,''
  arXiv:0804.0425 [astro-ph].

\bibitem{Beltran:2008ei}
  M.~Beltran,
  ``Isocurvature, non-gaussianity and the curvaton model,''
  arXiv:0804.1097 [astro-ph].

\bibitem{Li:2008jn}
  M.~Li, C.~Lin, T.~Wang and Y.~Wang,
  ``Non-Gaussianity, Isocurvature Perturbation, Gravitational Waves and a No-Go
  Theorem for Isocurvaton,''
  arXiv:0805.1299 [astro-ph].

\bibitem{Li:2008fm}
  S.~Li, Y.~F.~Cai and Y.~S.~Piao,
  ``DBI-Curvaton,''
  arXiv:0806.2363 [hep-ph].

\bibitem{Huang:2008qf}
  Q.~G.~Huang,
  ``Spectral Index in Curvaton Scenario,''
  arXiv:0807.0050 [hep-th].


\bibitem{Starobinsky:1986fxa}
  A.~A.~Starobinsky,
  ``Multicomponent de Sitter (Inflationary) Stages and the Generation of
  Perturbations,''
  JETP Lett.\  {\bf 42} (1985) 152.

\bibitem{Sasaki:1995aw}
  M.~Sasaki and E.~D.~Stewart,
  ``A General Analytic Formula For The Spectral Index Of The Density
  Perturbations Produced During Inflation,''
  Prog.\ Theor.\ Phys.\  {\bf 95}, 71 (1996)
  [arXiv:astro-ph/9507001].

\bibitem{Lyth:2005fi}
  D.~H.~Lyth and Y.~Rodriguez,
  ``The inflationary prediction for primordial non-gaussianity,''
  Phys.\ Rev.\ Lett.\  {\bf 95}, 121302 (2005)
  [arXiv:astro-ph/0504045].


\bibitem{Lyth:2004gb}
  D.~H.~Lyth, K.~A.~Malik and M.~Sasaki,
  ``A general proof of the conservation of the curvature perturbation,''
  JCAP {\bf 0505}, 004 (2005)
  [arXiv:astro-ph/0411220].




\bibitem{Lyth:2003dt}
  D.~H.~Lyth,
  ``Can the curvaton paradigm accommodate a low inflation scale,''
  Phys.\ Lett.\  B {\bf 579}, 239 (2004)
  [arXiv:hep-th/0308110].


\bibitem{Enqvist:2005pg}
  K.~Enqvist and S.~Nurmi,
  ``Non-gaussianity in curvaton models with nearly quadratic potential,''
  JCAP {\bf 0510}, 013 (2005)
  [arXiv:astro-ph/0508573].


\bibitem{Seery:2005gb}
  D.~Seery and J.~E.~Lidsey,
  ``Primordial non-gaussianities from multiple-field inflation,''
  JCAP {\bf 0509}, 011 (2005)
  [arXiv:astro-ph/0506056].

\bibitem{Lyth:1996im}
  D.~H.~Lyth,
  ``What would we learn by detecting a gravitational wave signal in the  cosmic
  microwave background anisotropy?,''
  Phys.\ Rev.\ Lett.\  {\bf 78}, 1861 (1997)
  [arXiv:hep-ph/9606387].



\bibitem{Huang:2007gk}
  Q.~G.~Huang,
  ``Weak gravity conjecture constraints on inflation,''
  JHEP {\bf 0705}, 096 (2007)
  [arXiv:hep-th/0703071].

\bibitem{Huang:2007qz}
  Q.~G.~Huang,
  ``Constraints on the spectral index for the inflation models in string
  landscape,''
  Phys.\ Rev.\  D {\bf 76}, 061303 (2007)
  [arXiv:0706.2215 [hep-th]].

\bibitem{Huang:2007st}
  Q.~G.~Huang,
  ``Weak Gravity Conjecture for the Effective Field Theories with N Species,''
  Phys.\ Rev.\  D {\bf 77}, 105029 (2008)
  [arXiv:0712.2859 [hep-th]].

\bibitem{Chen:2006hs}
  X.~Chen, S.~Sarangi, S.~H.~Henry Tye and J.~Xu,
  ``Is brane inflation eternal?,''
  JCAP {\bf 0611}, 015 (2006)
  [arXiv:hep-th/0608082].


\bibitem{Baumann:2006cd}
  D.~Baumann and L.~McAllister,
  ``A microscopic limit on gravitational waves from D-brane inflation,''
  Phys.\ Rev.\  D {\bf 75}, 123508 (2007)
  [arXiv:hep-th/0610285].


\bibitem{McAllister:2007bg}
  L.~McAllister and E.~Silverstein,
  ``String Cosmology: A Review,''
  Gen.\ Rel.\ Grav.\  {\bf 40}, 565 (2008)
  [arXiv:0710.2951 [hep-th]].



\bibitem{Bunch:1978yq}
  T.~S.~Bunch and P.~C.~W.~Davies,
  ``Quantum Field Theory In De Sitter Space: Renormalization By Point
  Splitting,''
  Proc.\ Roy.\ Soc.\ Lond.\  A {\bf 360} (1978) 117.


\bibitem{Vilenkin:1982wt}
  A.~Vilenkin and L.~H.~Ford,
  ``Gravitational Effects Upon Cosmological Phase Transitions,''
  Phys.\ Rev.\  D {\bf 26}, 1231 (1982).

\bibitem{Linde:1982uu}
  A.~D.~Linde,
  ``Scalar Field Fluctuations In Expanding Universe And The New Inflationary
  Universe Scenario,''
  Phys.\ Lett.\  B {\bf 116}, 335 (1982).


\bibitem{Jungman:1995df}
  G.~Jungman, M.~Kamionkowski and K.~Griest,
  ``Supersymmetric dark matter,''
  Phys.\ Rept.\  {\bf 267}, 195 (1996).



\bibitem{Huang:2007ek}
  Q.~G.~Huang,
  ``Simplified Chain Inflation,''
  JCAP {\bf 0705}, 009 (2007)
  [arXiv:0704.2835 [hep-th]].

\bibitem{Huang:2008jr}
  Q.~G.~Huang and S.~H.~Tye,
  ``The Cosmological Constant Problem and Inflation in the String Landscape,''
  arXiv:0803.0663 [hep-th].

\bibitem{Chialva:2008zw}
  D.~Chialva and U.~H.~Danielsson,
  ``Chain inflation revisited,''
  arXiv:0804.2846 [hep-th].


\bibitem{Silverstein:2008sg}
  E.~Silverstein and A.~Westphal,
  ``Monodromy in the CMB: Gravity Waves and String Inflation,''
  arXiv:0803.3085 [hep-th].


\bibitem{Dine:1986vd}
  M.~Dine and N.~Seiberg,
  ``Nonrenormalization Theorems in Superstring Theory,''
  Phys.\ Rev.\ Lett.\  {\bf 57}, 2625 (1986).


\bibitem{Dimopoulos:2005ac}
  S.~Dimopoulos, S.~Kachru, J.~McGreevy and J.~G.~Wacker,
  ``N-flation,''
  arXiv:hep-th/0507205.


\bibitem{Easther:2005zr}
  R.~Easther and L.~McAllister,
  ``Random matrices and the spectrum of N-flation,''
  JCAP {\bf 0605}, 018 (2006)
  [arXiv:hep-th/0512102].


\bibitem{Kachru:2003aw}
  S.~Kachru, R.~Kallosh, A.~Linde and S.~P.~Trivedi,
  ``De Sitter vacua in string theory,''
  Phys.\ Rev.\  D {\bf 68}, 046005 (2003)
  [arXiv:hep-th/0301240].

\bibitem{Svrcek:2006yi}
  P.~Svrcek and E.~Witten,
  ``Axions in string theory,''
  JHEP {\bf 0606}, 051 (2006)
  [arXiv:hep-th/0605206].

\bibitem{Dvali:1998pa}
  G.~R.~Dvali and S.~H.~H.~Tye,
  ``Brane inflation,''
  Phys.\ Lett.\  B {\bf 450}, 72 (1999)
  [arXiv:hep-ph/9812483].

\bibitem{Kachru:2003sx}
  S.~Kachru, R.~Kallosh, A.~Linde, J.~M.~Maldacena, L.~P.~McAllister and S.~P.~Trivedi,
  ``Towards inflation in string theory,''
  JCAP {\bf 0310}, 013 (2003)
  [arXiv:hep-th/0308055].

\bibitem{ArkaniHamed:2003uz}
  N.~Arkani-Hamed, P.~Creminelli, S.~Mukohyama and M.~Zaldarriaga,
  ``Ghost inflation,''
  JCAP {\bf 0404}, 001 (2004)
  [arXiv:hep-th/0312100].

\bibitem{Chen:2006nt}
  X.~Chen, M.~x.~Huang, S.~Kachru and G.~Shiu,
  ``Observational signatures and non-Gaussianities of general single field
  inflation,''
  JCAP {\bf 0701}, 002 (2007)
  [arXiv:hep-th/0605045].

\bibitem{Bean:2008ga}
  R.~Bean, D.~J.~H.~Chung and G.~Geshnizjani,
  ``Reconstructing a general inflationary action,''
  arXiv:0801.0742 [astro-ph].

\bibitem{Li:2008qc}
  M.~Li, T.~Wang and Y.~Wang,
  ``General Single Field Inflation with Large Positive Non-Gaussianity,''
  JCAP {\bf 0803}, 028 (2008)
  [arXiv:0801.0040 [astro-ph]].


\bibitem{Chen:2007gd}
  B.~Chen, Y.~Wang and W.~Xue,
  ``Inflationary NonGaussianity from Thermal Fluctuations,''
  JCAP {\bf 0805}, 014 (2008)
  [arXiv:0712.2345 [hep-th]].

\bibitem{Matsuda:2008hx}
  T.~Matsuda,
  ``Modulated Inflation,''
  arXiv:0801.2648 [hep-ph].

\bibitem{Matsuda:2008fk}
  T.~Matsuda,
  ``Running spectral index from shooting-star moduli,''
  JHEP {\bf 0802}, 099 (2008)
  [arXiv:0802.3573 [hep-th]].

\bibitem{Gao:2008dt}
  X.~Gao,
  ``Primordial Non-Gaussianities of General Multiple Field Inflation,''
  arXiv:0804.1055 [astro-ph].

\bibitem{Li:2008qv}
  S.~W.~Li and W.~Xue,
  ``Revisiting non-Gaussianity of multiple-field inflation from the field
  equation,''
  arXiv:0804.0574 [astro-ph].

\bibitem{Xue:2008mk}
  W.~Xue and B.~Chen,
  ``$\alpha$-vacuum and inflationary bispectrum,''
  arXiv:0806.4109 [hep-th].







\end{thebibliography}
\end{document}